\documentclass[fleqn,twoside]{article}
\usepackage{espcrc2,epsfig}

\usepackage{graphicx}
\usepackage[figuresright]{rotating}

\newcommand{\AmS}{{\protect\the\textfont2
  A\kern-.1667em\lower.5ex\hbox{M}\kern-.125emS}}

\title{
Recent CLEO Results on Tau Hadronic Decays
}

\author{J.E. Duboscq\address{ Wilson Laboratory, Cornell University, Ithaca NY 14850, USA}
        \thanks{I am thankful for the fine efforts of the CESR staff for
         creating the many collisions which make this work possible. I would also like to
          thank the National Science Foundation for its essential support. The conference organizers
           merit many thanks for their very enjoyable conference.  
 Presented at the 8th International Workshop on Tau-Lepton Physics, Nara, Japan, 
         Sept 2004. To appear in Nuclear Physics B - Proceedings Supplements.
}
          for the CLEO collaboration
          }

\begin{document}

\begin{abstract}
I outline recent results using CLEO3 data involving the decay of the $\tau$ to three 
 charged hadrons and a neutrino, as well as an investigation of the structure of the
  decay $\tau \to KK\pi \nu$ and the Wess-Zumino term in this decay.
\vspace{1pc}
 \end{abstract}

\maketitle

\section{The Decay of the $\tau$ to Three Charged Hadrons and a neutrino}

When the $\tau$ lepton decays to 3 charged hadrons and a neutrino, the predominant
 decay mode involves three charged pions. Occasionally this decay will  include
 one, two or three kaons in the final state. Each of these modes is useful for understanding
  different physics topics. The decay $\tau^- \to K^- \pi^+\pi^- \nu$ is important in 
   extracting the strange spectral function as well the strange quark mass and the
    CKM matrix element $V_{us}$. The decay $\tau^- \to K^- K^+\pi^- \nu$
allows one to probe the Wess Zumino term in the effective lagrangian. Finally, the
 decay $\tau \to K^-K^+K^-\nu$ is as yet unobserved.
 
 In this analysis we use a subset of almost 3 million $\tau$ pairs produced at the CESR
  $e^+e^-$ collider at or near the $\Upsilon(4S)$ resonance, analyzed with the CLEO3~\cite{cleo3det}
   detector. Since the predominant decay of the $\tau$ to 3 charged hadrons usually involves
    pions, we use the CLEO3 RICH detector to identify Kaons with high efficiency while 
     rejecting pion fakes. This is combined with the detector's $dE/dx$ capabilities to 
      identify pions and kaons with efficiencies in the $\approx 90 \%$ region and
       reject fakes at the $\approx 3\%$ level up to momenta of almost 2 GeV/c.
       These efficiencies and fake rates are extracted from data using the decay
        chain $D^*\to D^0\pi, D^0\to K\pi$.\footnote{ As a consistency , we looked for wrong sign
         kaons in the decay $\tau^- \to K^+ \pi^-\pi^+ \nu$, and found that data and
          Monte Carlo estimations agreed.}
  
  At CLEO near the $\Upsilon(4S)$ resonance, $\tau$'s are produced back to back with
   substantial momentum. This allows us to select $\tau$ events by looking for events
    with 1 track recoiling against 3 tracks, with the event hemispheres identified using
     the event thrust axis.
 The single track hemisphere is required to be consistent with either a $\tau$ decay to
  an electron, muon, single pion, or $\rho$ with an unseen neutrino.
  Events with extra showers are rejected to cut down on the feed-through from 
   $\tau \to 3h \pi^0 \nu$ events. Missing momentum and visible energy
    cuts reduce 2 photon fusion background events. Also, events consistent with
    the production of a $K^0_S$ are rejected in the $K\pi\pi$ mode.
    Events are simulated using KORALB, JETSET and GEANT, while particle
     identification efficiencies are taken from data as noted above.
     
 \begin{table}
\caption{ Candidate event yields, estimated $\tau$ and $q\overline{q}$ backgrounds, and efficiencies for $\tau \to 3h\nu$ }
\begin{center}
\begin{tabular}{|c|c|c|c|c|} \hline
         Mode	& Data  & $\tau$ bgd & $q\overline{q}$ bgd &$\epsilon (\%)$ \\ \hline 
$\pi\pi\pi$ & $43543$ & $3207\pm57$ & $152\pm12$ & $10.27 \pm 0.08$ \\ \hline
$K\pi\pi$ & $3454$ & $1475\pm38$ & $57\pm8$ & $11.63\pm0.12$ \\ \hline
$KK\pi$ & $932$ & $86\pm9$ & $19\pm4$ & $12.48\pm 0.11$ \\ \hline
$KKK$ & $ 12 $ & $ 4\pm2 $ & $0.4\pm 0.6$ & $9.43 \pm 0.10$ \\ \hline
\end{tabular}
\end{center}
\label{table:1}
\end{table}

\begin{figure}
\psfig{figure=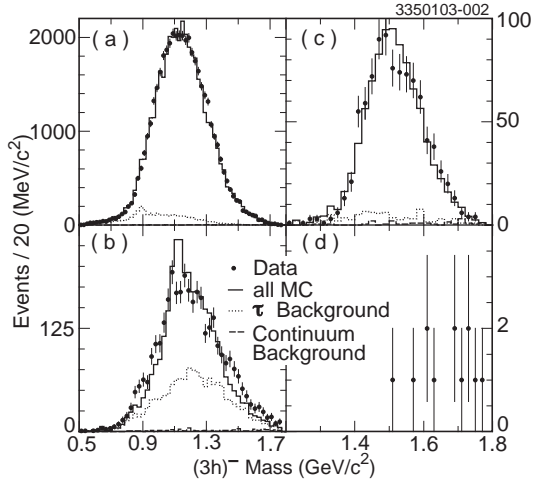, ,scale=0.40}
\vskip -1cm
\caption{ Three hadron masses from data (points). Tau pair Monte Carlo yields are shown as short dashes,
${q\overline{q}}$ Monte Carlo as long dashes; the sum of the background and signal MC is the solid histogram. 
 (a) $\pi\pi\pi$ mass, (b) $K\pi\pi$ mass, (c) $KK\pi$ mass, (d) $KKK$ mass }
\label{fig:fig1}
\end{figure}

\begin{figure}
\psfig{figure=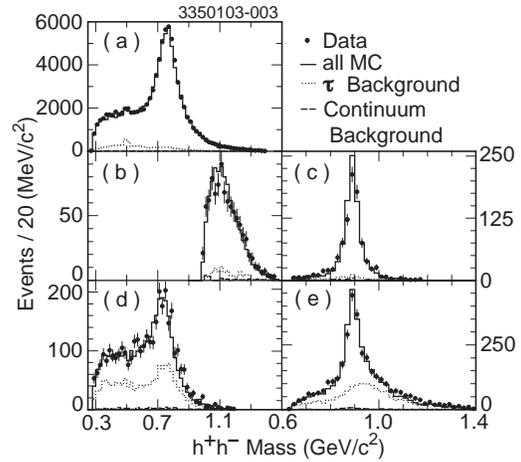, ,scale=0.40}
\vskip -1cm
\caption{ Two body substructure of the 3 body mass samples for oppositely charged hadrons. Data are shown as points, Tau pair Monte Carlo yields are shown as short dashes,
${q\overline{q}}$ Monte Carlo as long dashes; the sum of the background and signal MC is the solid histogram.
 (a) $\pi\pi$ mass in $\pi\pi\pi$ (2 entries per event) ,
 (b) $KKi$ mass in $KK\pi$ ,
 (c) $K\pi$ mass in $KK\pi$ ,
 (d) $\pi\pi$ mass in $K\pi\pi$ ,
  (e) $K\pi$ mass in $K\pi\pi$
  }
\label{fig:fig2}
\end{figure}

The breakdown of the analysis results are shown in Table~\ref{table:1}. The largest $\tau$ backgrounds
 are feedthroughs from $\tau \to 3h (\pi^0)\nu$ modes. The feed-across is determined
  from the Monte Carlo, except in the $KKK$ signal mode in which the data is used.    The resulting three body mass plots are shown in Fig~\ref{fig:fig1}.
   For each of these channels the two body mass structure is shown in Fig~\ref{fig:fig2}.
   These plots show very good agreement between data and Monte Carlo. The tunings used
    for the $\pi\pi\pi$ and $K\pi\pi$ components are those reported at TAU02~\cite{tau02}. The channel
     $KK\pi$ had its substructure tuned to the data. This required decreasing the 
      $K^*$ contribution, increasing the $\rho'$, and removing the $\rho''$ component entirely.
      
The dominant systematic error in this study comes from particle identification
 efficiency ($3\%$) and particle identification fakes (ranging from
  $0.1\%$ to $12\% $.) These were determined using the both the $D^*$ sample
   referred to above, as well as a search for the  wrong sign decay $\tau^- \to K^+ \pi^-\pi^- \nu$.
   A systematic error of $2\%$ was estimated for the tuning of
    the $KK\pi$ substructure.
    
    Final branching ratio results are found to be:
    \begin{eqnarray*}
    B(\tau^- \to \pi^-\pi^+\pi^- \nu ) & = & 9.13 \pm 0.05 \pm 0.46 \% \\
    B(\tau^- \to K^-\pi^+\pi^- \nu ) & = & 0.384 \pm 0.014 \pm 0.038 \% \\
     B(\tau^- \to K^-K^+\pi^- \nu ) & = & 0.155 \pm 0.006 \pm 0.009 \% \\
    B(\tau^- \to K^-K^+K^- \nu ) & < & 3.7 \times 10^{-5}\, @ \, 90 \% CL \\
    \end{eqnarray*}
    
    The $\pi\pi\pi$ result is the first in which all three pions are explicitly
     identified. The $K\pi\pi$ result is consistent with the previous OPAL result of 
      $0.360 \pm 0.082\pm 0.048 \%$ ~\cite{OPALkpipi}, and the CLEO2~\cite{cleo2kpipi}  result
       of $0.346\pm0.023\pm0.056\%$. The result is  higher
        than the ALEPH reported value~\cite{ALEPHkpipi} of $0.214\pm 0.037 \pm 0.029\%$.
        The $KK\pi$ result is the most precise currently available, while the $KKK$ limit
         is the most stringent.
    This work has been published in ~\cite{cleo3h}

\section{ The Structure of $\tau \to KK\pi\nu$ and the Wess Zumino Anomaly }

  In the simplest picture of $\tau$ decays, vector (axial) currents produce an even (odd) number of 
pseudoscalars in the final state.
 The Wess Zumino anomaly term in the effective Lagrangian allows for a parity flip and can
  cause a violation of this simple rule.
  The golden mode for this mechanism $\tau \to \eta \pi \pi^0 \nu$ contains no axial term
   and has been previously observed by CLEO (~\cite{cleoetapipi}).
   The decay $\tau \to KK \pi \nu$ is expected to have both axial and vector contributions.
    Extracting the vector component allows one to examine the WZ term in this decay.
    
    The $\tau$ decay to 3 hadrons and a neutrino can be expressed as a product of a leptonic
     current and a hadronic current~\cite{kuhn}. The hadronic current can be expanded into a sum
      over 4 form factors ($F_i$):
      $ J = \Sigma f_i(q_1,q_2,q_3) F_i(s_1,s_2, Q)$. In this expression, the $q_i$ are the individual hadron
       four momenta, the $s_i$ are the pairwise hadron momenta, and $Q$ is the total four momentum of
        the hadronic system.
       The $f_i$ are kinematic terms. The $F_1$ and $F_2$ are the axial vector terms. The $F_3$ 
        term is the W-Z vector term and its corresponding kinematic term $f_3$ is 
         $i \epsilon^{\alpha\beta\gamma}q_{1\alpha}q_{2\beta}q_{3\gamma}$. This $\epsilon$
          invokes the parity flip that allows the WZ mechanism to operate. (The $F_4$ term 
           corresponds to the negligible scalar current.)
           
   In order to extract these terms, we integrate over the unobserved neutrino direction.
    The two remaining Euler angles are kinematically determined at CLEO. With this
     it turns out that the differential decay rate can be expressed as:
     $d\Gamma(\tau\to K K \pi)/dQ^2ds_1ds_2 \propto W_A(F_1,F_2) + W_B(F_3) $.
     The $W_B$ term expresses the strength of the W-Z anomaly. Note that there
      is no interference between the $W_A$ and $W_B$ terms, and  the extraction
       of these terms is possible by using only subcomponent masses and the $Q^2$ evolution of
        the decay.
        
  In order to actually do this fit, we need to assume some model for the physics involved. We have
   used the following modification of the model in \cite{Decker,Finkemeir}:
   For the $F_1$ term, we allow the decay to proceed through $a_1 \to \rho^{(')} \pi, \rho^{(')} \to KK$. The $F_1$ term 
   thus is proportional to $BW_{a1}(Q^2) \times ( BW_\rho(s_2)+\beta_\rho BW_\rho'(s_2) )$, where BW denotes a Breit Wigner function.
   For the $F_2$ term, the decay proceeds through $a_1 \to K^*K, K^*\to K\pi$. The $F_2$ term is proportional to
   $R_F BW_{a_1}(Q^2) BW_{K^*}(s_1)$.
   For the W-Z term, we parameterize the decay as occurring either through the channels
    $\rho^{(','')} \to K^*K, K^*\to K \pi$, or through $\rho^{(','')} 
     \to \omega \pi, \omega \to KK$. This leads to
     $F_3 \propto R_B^{1/2} ( BW_\rho(Q^2)+\lambda BW_{\rho'}(Q^2) + \delta BW_{\rho''}(Q^2) )
      \times ( BW_\omega(s_2) + \alpha BW_{K^*}(s_1)) $
      
  In the above there are 5 real parameters to fit: $R_F$, $R_B$, $\lambda$, $\delta$, and
   $\alpha$. Note that these fits are only used to model the decay and do not necessarily
    correspond to the correct physics. These five parameters are extracted from fits
     to the $KK\pi$, $K\pi$ and $KK$ mass plots.
    
 The data used for this analysis corresponds to a superset of that of the previous analysis, 
  comprising some 7 million $\tau$ pairs at CLEO3. The cuts used are the same as
   in the $\tau \to 3h\nu$ analysis above, and result in 2255 signal events with
    an estimated $256\pm16\pm46$ background events. The absolute branching ratio
     obtained with this sample is consistent with that given above.
     
    To extract the values of the five parameters we perform an unbinned extended maximum
     likelihood fit including a background term. The probability distribution function that
      is fit is a product of the individual PDFs for $KK\pi$, $KK$, $K\pi$. We use the best
       known values of the parameters for each Breit Wigner term.

\begin{figure}[htb]
\psfig{figure=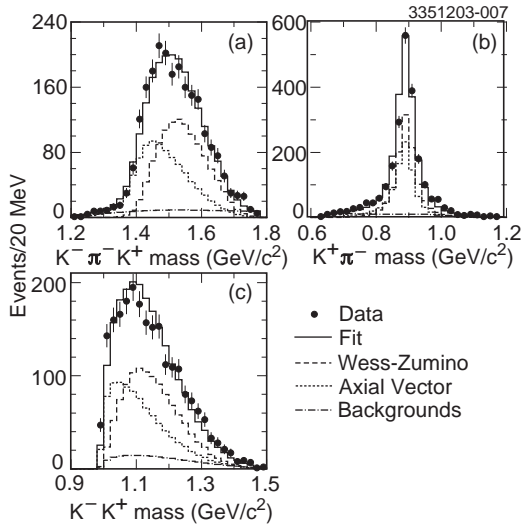, ,scale=0.40}
\vskip -1cm
\caption{ Fit projections (solid lines), data (points), and contributions from the W-Z term (dashed) and the axial term (dotted), as well as all backgrounds (dot-dashed) for the $KK\pi$ W-Z analysis.
 (a)$ KK\pi$ mass projection, 
 (b) $K^+\pi^-$ mass projection, 
 (c) $K^-K^+$ mass projection
  }
\label{fig:fig3}
\end{figure}

 Fig~\ref{fig:fig3} shows the results of the fit for $KK\pi$, $K\pi$ and $KK$. Also shown on the plots
 are the WZ (vector) component and the axial component. It is clear that the W-Z term is
  prominent in the fit - the result is that the partial width from W-Z is:
  $\Gamma_{WZ}/\Gamma_{Tot} = 55 \pm 8.4 \pm 4.9 \% $. 
  The best fit  parameters are given in Table~\ref{table:2}. 
  
 \begin{table}[ht]
\caption{ W-Z Fit results }
\begin{tabular}{|c|c|} \hline
$\alpha$ & $0.471 \pm 0.060 \pm 0.034$ \\ \hline
$\lambda$ & $0.314\pm0.073\pm 0.080 $ \\ \hline
$\delta$ & $0.101 \pm 0.020 \pm 0.156 $ \\ \hline
$ R_B $ & $3.23\pm0.26\pm1.90$ \\ \hline
$ R_F $ & $0.98 \pm 0.15 \pm 0.36 $ \\ \hline \hline 
$\Gamma_{WZ}/\Gamma_{Tot} $ & $ 55 \pm 8.4 \pm 4.9 \% $ \\ \hline
\end{tabular}
\label{table:2}
\end{table}

The substructure of this fit can be expressed in the context of our modified Kuhn \& Mirkes model. The
 decay is found to be dominated by the $K^*K$ mode, with approximately equal W-Z and axial components.
 Explicitly,
 \begin{eqnarray*}
 R_{WZ}^{\omega\pi} &  =  & 3.4 \pm 0.9 \pm 1.0 \%  \\
  R_{Axial}^{\rho^{(')}\pi} &  =  & 2.50 \pm 0.8 \pm 0.4 \% \\
 R_{WZ}^{K^*K}  & = &  60.8 \pm 8.5 \pm 6.0 \%  \\
  R_{Axial}^{K^*K} & = & 46.8 \pm 8.4 \pm 5.2\% 
 \end{eqnarray*}
 These fractions do not add up to $100\%$ because of interference terms in the intermediate states.
 
 We extract the value of the branching fraction $B(a_1 \to K^*K) = 2.2 \pm 0.5 \%$ 
and this is found to be consistent with expectations from the CLEO analysis for the
 decay $\tau \to 3\pi^0 \nu$.
 Note that the axial component determined directly here is much smaller than that
  which has been found by Aleph's CVC estimate ( $94 ^{+6}_{-8} \%$ )~\cite{ALEPHaxial}
    using data from the DM1 and DM2 collaborations. Note that our fit does have a floating
     value of $R_B$ which might contribute to the observed difference.
This work has been published in ~\cite{cleowz}.
     
  \section{Summary}
  In this talk, we have presented the first direct measurement of $B(\tau- \to \pi^- \pi^+ \pi^- \nu_\tau)$.
  We have also presented a measurement of $B(\tau- \to K^- \pi^+ \pi^- \nu_\tau)$ consistent with
   previous work from OPAL and CLEO, but higher than the result found by ALEPH.
   We've set the most stringent limit on the branching fraction of the decay of a $\tau$ to 3 kaons and
    a neutrino, and have presented the best precision on $B(\tau \to K^-K^+\pi^- \nu_\tau)$.
     We have also presented the first study of the WZ anomaly and the axial component of
      the decay $\tau^- \to K^-K^+\pi^- \nu_\tau$.


\begin{thebibliography}{9}
\bibitem{oldcleo2result} R. Ammar {\it et al} (CLEO Collaboration ),Phys Rev D {\bf 49} ,5701 (1994)
\bibitem{cleodet} Y.Kubota {\it et al} (CLEO Collaboration), Nucl. Instrum. Methods Phys. Res., Sect. A {\bf 320}, 66 (1992) ; T. Hill, Nucl. Instrum. Methods Phys. Res. Sect. A {\bf 418}, 32 (1998).
\bibitem{cleo3det} G. Viehhauser, {\it CLEO III Operation},Nucl. Instrum. Methods A {\bf 462}, 146 (2001).
\bibitem{cleo2kpipi} S.J. Richichi {\it et al} (CLEO Collaboration) , Phys. Rev. D 60, 112002 (1999) 
\bibitem{tau02} A. Weinstein, hep-ex/0210058 
\bibitem{OPALkpipi} G.Abbiendi {\it et al} (OPAL Collaboration) Euro. Phys. Journal C 13, 197 (2000.)
\bibitem{ALEPHkpipi} R. Barate {\it et al} (ALEPH Collaboration), Euro.Phys. Journal C1, 65 (1998)
\bibitem{cleo3h}  R. Briere {\it et al} (CLEO Collaboration),  Phys. Rev. Lett. 90, 181802 (2003).
\bibitem{cleoetapipi} M. Artuso {\it et al} (CLEO Collaboration), Phys.Rev.Lett 69, 3278 (1992)
\bibitem{kuhn} Kuhn, Mirkes, Z.PhysC56, 661(1992).
\bibitem{Decker} Decker {\it et al}, Z.Phys.C58, 445(1996).
\bibitem{Finkemeir} Finkemeir \& Mirkes, Z,Phys.C69, 24(1996).
\bibitem{ALEPHaxial}  R. Barate {\it et al} (ALEPH Collaboration), Euro. Phys. Journal C 11 (1999) 599-618 
\bibitem{cleowz} T.E. Coan {\it et al} (CLEO Collaboration) , Phys. Rev. Lett. 92, 232001 (2004)
\end{thebibliography}
\end{document}